\documentclass[useAMS,usenatbib]{mn2e}

\usepackage{graphicx}
\usepackage{lscape}
\usepackage{amsmath}
\usepackage{color}
\usepackage{placeins}
\newcommand{\msolar}{\mbox{\,M$_\odot$}}        
\newcommand{\kms}{\mbox{km\,s$^{-1}$}}  

\title [The Gas Properties of the W3 GMC]{The Gas Properties of the W3 GMC: A HARP study}
\author[D. Polychroni, T.J.T. Moore \& J. Allsopp]{D. Polychroni$^{1,2}$\thanks{E-mail:
danae.polychroni@ifsi-roma.inaf.it (DP);}, T.J.T. Moore$^{1}$ \& J. Allsopp$^{1}$ \\
${^1}$Astrophysics Research Institute, Liverpool John Moores University, UK \\
${^2}$ Istituto Nationale di Astrofisica, Istituto di Astrofisica e Planetologia Spaziali, Roma, Italy}

\begin{document}

\date{Accepted 2012 February 20. Received 2012 February 18; in original form 2011 July 19}

\pagerange{\pageref{firstpage}--\pageref{lastpage}} \pubyear{2011}

\maketitle

\label{firstpage}

\begin{abstract}

We present $^{12}$CO, $^{13}$CO and C$^{18}$O J=3$\to$2 maps of the W3 GMC made at the James Clerk Maxwell Telescope. We combine these observations with Five Colleges Radio Astronomy Observatory CO J=1$\to$0 data to produce the first map of molecular-gas temperatures across a GMC and the most accurate determination of the mass distribution in W3 yet obtained.  We measure excitation temperatures in the part of the cloud dominated by triggered star formation (the High Density Layer, HDL) of 15-30\,K, while in the rest of the cloud, which is relatively unaffected by triggering (Low Density Layer, LDL), the excitation temperature is generally less than 12\,K. We identify a temperature gradient in the HDL which we associate with an age sequence in the embedded massive star-forming regions. We measure the mass of the cloud to be 4.4$\pm$0.4$\times$10$^{5}\msolar$, in agreement with previous estimates. Existing sub-mm continuum data are used to derive the fraction of gas mass in dense clumps as a function of position in the cloud.  This fraction, which we interpret as a Clump Formation Efficiency (CFE), is significantly enhanced across the HDL, probably due to the triggering. Finally, we measure the 3D rms Mach Number, $\mathcal{M}$, as a function of position and find a correlation between $\mathcal{M}$ and the CFE within the HDL only. This correlation is interpreted as due to feedback from the newly-formed stars and a change in its slope between the three main star-forming regions is construed as another evolutionary effect. We conclude that triggering has affected the star-formation process in the W3 GMC primarily by creating additional dense structures that can collapse into stars. Any traces of changes in CFE due to additional turbulence have since been overruled by the feedback effects of the star-forming process itself.
\end{abstract}


\section{Introduction}

One of the most important current questions in the field of star formation concerns the effect that environment and, especially, feedback may have on the star-formation process, in particular the stellar initial mass function (IMF) and the star-formation rate (SFR) or efficiency (SFE).  Stars appear to form in two main modes.  Spontaneous star formation is the predicted result of the naturally turbulent molecular-cloud environment (see e.g. \citealt{2004RvMP...76..125M}; \citealt{2004ASPC..322..299K}; \citealt{2006ApJ...648.1052H}; \citealt{2002ApJ...576..870P}) and is expected to produce a low background SFR.  Triggered star formation, on the other hand, is an increase in SFR or SFE due to the effects of a mechanical interaction on molecular cloud gas, usually caused by the winds, radiation or expanding \textsc{Hii} regions associated with massive stars (see e.g. \citealt{1998ASPC..148..150E}; \citealt{2005A&A...433..565D}; \citealt{2011MNRAS.418.2121G}).

There are two main ways in which triggering might increase the SFE locally. The first is by creating new star-forming structures (i.e.\ dense cores), in addition to those forming spontaneously in the turbulent molecular cloud gas. This mode is most closely described by the {\em collect-and-collapse} mechanism (\citealt{1977ApJ...214..725E}, \citealt{1994MNRAS.268..291W}) in which an expanding dense shell, driven into a cloud by winds or thermal expansion, becomes gravitationally unstable and fragments to form dense, star-forming clumps.  The second mode works by increasing the probability that pre-existing dense ``cores" will collapse to form stars.
This would usually require an increase in the ambient pressure, either from the passage of a (shock) wave through clumpy cloud gas, or when a core is overtaken by an ionisation front (IF).  The latter mechanism is described by the radiatively-driven implosion (RDI) model (\citealp{1980SSRv...27..275K}; \citealp{1989ApJ...346..735B}; \citealp{1989ApJ...342L..87S}; \citealp{1990ApJ...354..529B}) 

\begin{figure}
\centering
\includegraphics[height=8.4cm]{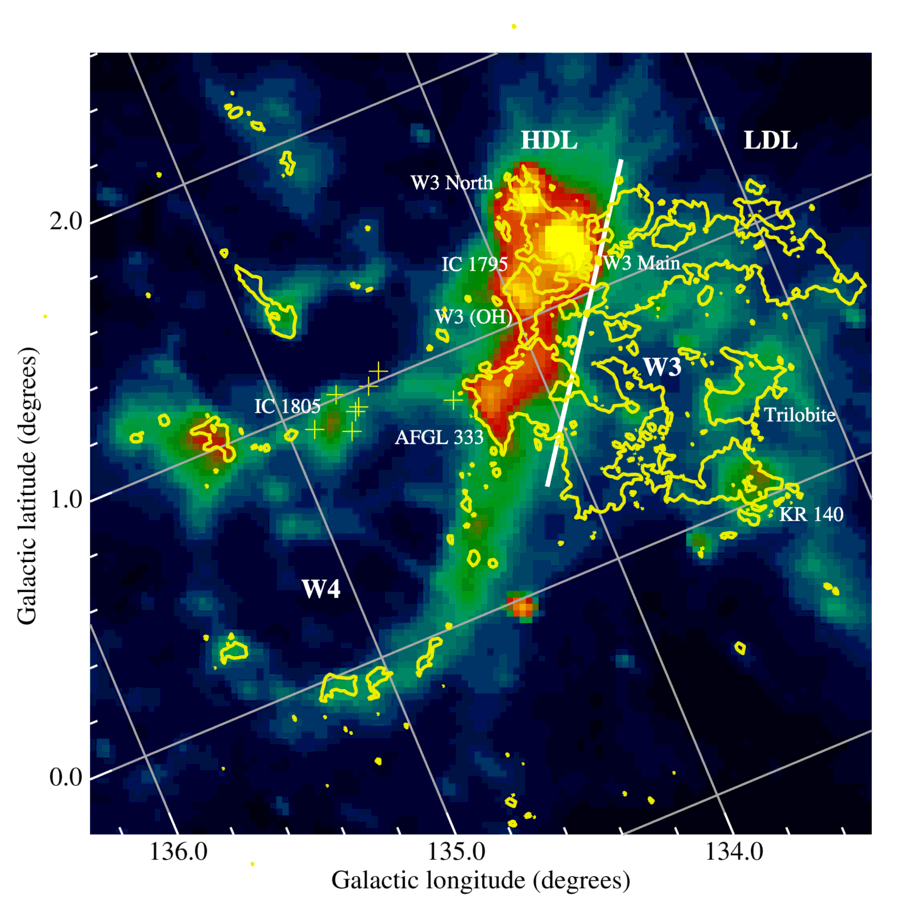}
\caption[small]{MSX image of the W3/4 region at 8$\,\mu$m. The black contours trace the $^{12}$CO J=1$\to$0 outline of the W3 GMC, as observed by the FCRAO. The white straight line marks the approximate boundary between the HDL and LDL regions. The yellow crosses mark the positions of the O stars in the IC 1805 cluster. Several regions of interest within the cloud are labeled.}
\label{fig:msxw4w3}
\end{figure}

A third potential type of mechanism for influencing the SFE would include any that affect the efficiency with which dense cores convert mass into stars.  This is the same as saying that the accretion history of already bound cores is affected by a change in the local environment. This might be caused by variations in local density and/or effective signal speed, which would alter the accretion rate.

The W3 Giant Molecular Cloud (GMC) offers a potential `laboratory' for constraining the mechanism by which feedback triggers new star formation and for quantifying increases in star-formation efficiency above the spontaneous background rate. Both modes of star formation appear to exist within W3, each dominating different parts of the cloud. Thought to be a prime example of triggered, sequential star formation (\citealp{1978ApJ...226L..39L}; \citealp{2005ApJ...620L..43O}), it  stands on the western side of the W4 chimney/superbubble whose expansion is driven by the winds of the IC 1805 OB association (Figure \ref{fig:msxw4w3}). This expansion is compressing the eastern side of the W3 GMC and has created a high density layer (HDL: \citealt{1978ApJ...226L..39L}) within which there are several prominent star-forming regions (i.e.\ W3\,Main, W3\,(OH) and AFGL\,333). The rest of the cloud seems so far to have been largely unaffected by this interaction and the star formation within it should be mainly spontaneous.  One notable exception is the KR\,140 \textsc{Hii} region, located in the far south-west corner of the cloud, this may be an example of the spontaneous formation of an isolated massive star which is now triggering new star formation in a surrounding shell (\citealt{2008MNRAS.385..995K}).

\citet{2007MNRAS.379..663M} surveyed two thirds of the W3 cloud, including all the HDL and the southern half of the remaining cloud, in the 850-$\umu$m continuum and detected 316 dense cores with masses above 13\msolar. Dividing the GMC crudely into the two zones, they found that a significantly greater fraction of the total gas mass is contained in dense, potentially star-forming structures in the HDL (25--37\%, depending on assumptions about the clump mass function, or CMF) compared to the diffuse cloud (5--13\%), but detected no difference in the CMF between the two sections of the cloud.  These results were interpreted as clear evidence of a collect-and-collapse type mechanism at work. However, this result was derived assuming a single excitation temperature (30\,K) everywhere in the molecular gas traced by CO J=1$\to$0. If the gas temperature were significantly higher in the HDL than in the remaining cloud, then the contrast in gas mass ratio between the two regions may be lower than this analysis suggests.

This paper presents new CO J=3$\to$2 emission line maps of the W3 GMC and an analysis of the physical excitation of the cloud molecular gas, in particular the distribution of excitation temperatures, using matching CO J=1$\to$0 data.  The fraction of gas mass in dense clumps is then estimated as a function of position using the existing 850-$\umu$m continuum results and is compared to the Mach number of the turbulence in the CO-traced gas. The paper is structured as follows: In Section \ref{sec:Obs} we detail the data reduction procedure for the CO J=3$\to$2 data and describe the CO J=1$\to$0 and SCUBA datasets.  In Section \ref{sec:Analysis} we describe our analysis and discuss the results. Finally, in Section \ref{sec:conc}, we present the conclusions of this study.

\section[]{Observations and Data Reduction.}
\label{sec:Obs}

\subsection{HARP Data}

\begin{figure*}
\includegraphics[width=\textwidth]{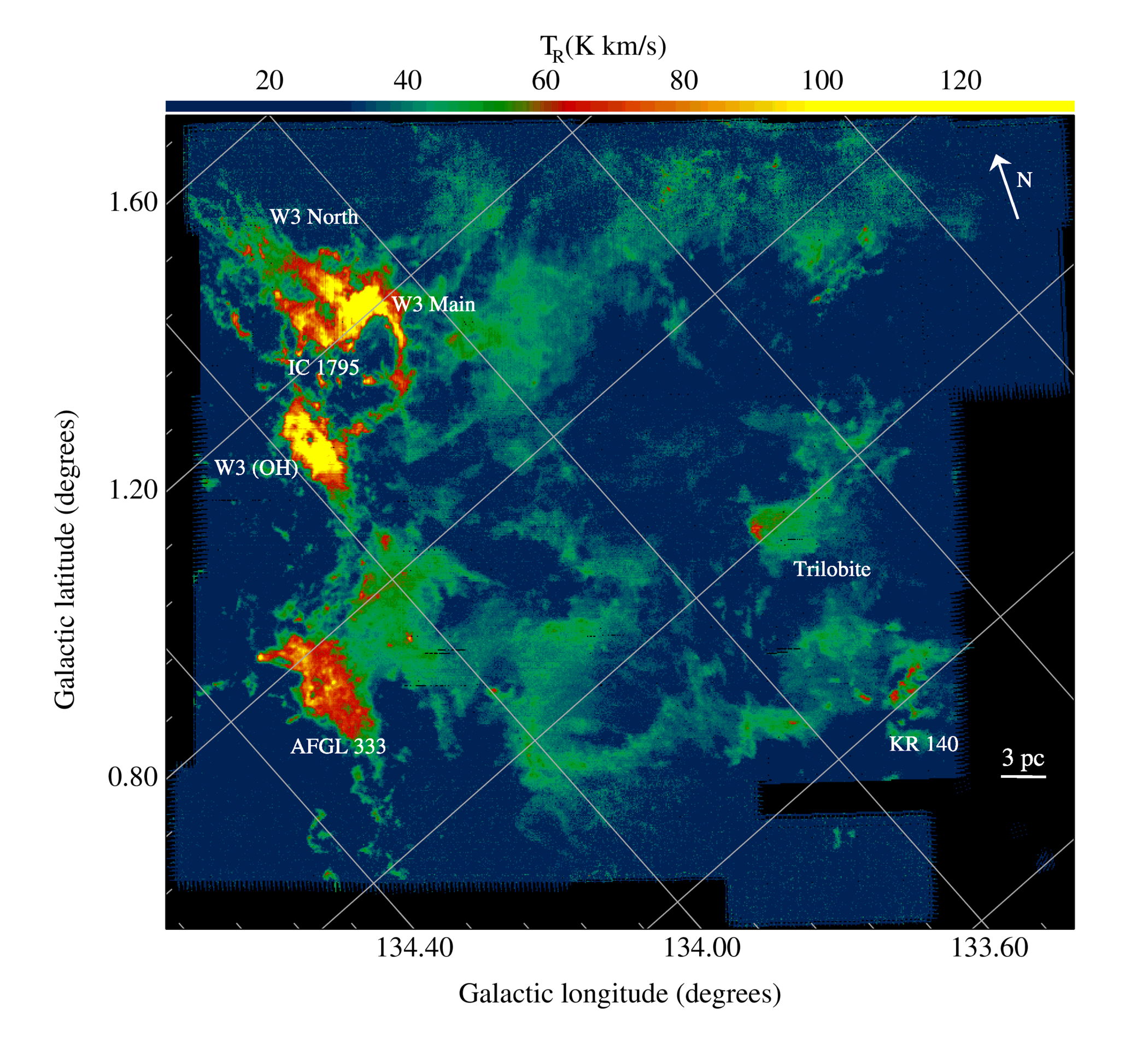}
\caption[small]{T$_{R}^{*}$ $^{12}$CO J=3$\to$2 emission from the W3 GMC, integrated in the range $-65<V_{\rm LSR}<-25$\,\kms, showing the total area surveyed in this transition. Several star-forming regions have been identified on the map. The grid denotes Galactic coordinates.}
\label{fig:12co}
\end{figure*}

The W3 GMC was mapped with the HARP array receiver on the James Clerk Maxwell Telescope (JCMT) on Mauna Kea, Hawaii. HARP is a 16-element focal-plane array that operates in the 325--375\,GHz regime \citep{2009MNRAS.399.1026B}. The 16 receptors have $\sim$14\,arcsec beams separated by 30 arcseconds, giving the instrument a 2-arcminute footprint. HARP is combined with the ACSIS digital autocorrelation spectrometer backend. 

Observations were made over three consecutive years (2006-2008) in $^{12}$CO, $^{13}$CO and C$^{18}$O J=3$\to$2 at 345.796, 330.540 and 329.278 GHz, respectively. All observations were taken in good weather with the sky opacity at 225\,GHz in the range $\tau_{225}<$0.08 (JCMT weather band categories 1 \& 2), using a bandwidth of 250\,MHz, giving a basic spectral resolution of 26\,m\,s$^{-1}$.

As the W3 GMC spans about 1$^{\rm o}$ on the sky, we split the cloud into 13 separate tiles of $20 \times 20$ arcminutes, each requiring about one hour to map. All tiles were observed using continuous raster scanning and pointing observations were made between each tile. We used a sample spacing of 7.5 arcseconds and the raster scan spacing was half an array footprint.  Scans were aligned at a position angle of 70$^{\rm o}$ to the Declination axis to match the known geometry of the cloud. We also observed small raster scan maps of CRL 618, CRL 2688 and W3 (OH) to calibrate the science maps. The calibration factors applied in the segments of the maps were between 1.12 - 1.5 for $^{12}$CO, 1.7 - 3.04 for $^{13}$CO and 1.02 - 1.3 for C$^{18}$O. The System Temperature varied between 233\,K and 283\,K for $^{12}$CO with a median value of 242\,K; 289\,K to 902\,K for $^{13}$CO with a median of 359\,K; and 298\,K to 340\,K for C$^{18}$O, with a median value of 324\,K. The mean pointing error was 2.43$\pm$0.33 arcseconds for all the observations. 

The observing procedure differed slightly over the years as new observing modes became available. In particular, the great majority of the $^{12}$CO and $^{13}$CO maps have been scanned only along one direction, as the ``basket weave" mode of orthogonal scanning was not available at the time. 
The C$^{18}$O map and later parts of the $^{13}$CO data were made using this mode. The velocity range of the cubes is $-120$\,km\,s$^{-1}$ to $+30$\,km\,s$^{-1}$.

The raw data cubes were filtered for spikes and, in $^{12}$CO, were binned by a factor of nine in the spectral axis to achieve a rms noise level of $\sim$0.7\,K  in a 0.23-km\,s$^{-1}$ wide channel. The $^{13}$CO and C$^{18}$O J=3$\to$2 maps were binned by a factor of 15 to obtain rms noise of $\sim$0.4\,K in a 0.39-km\,s$^{-1}$ wide channel. The maps were spatially regridded with pixel resolution of 7.7 arcseconds and the spectral baselines were removed. 

\subsection{FCRAO Data}

\begin{figure*}
\centering
\includegraphics[width=6.3in]{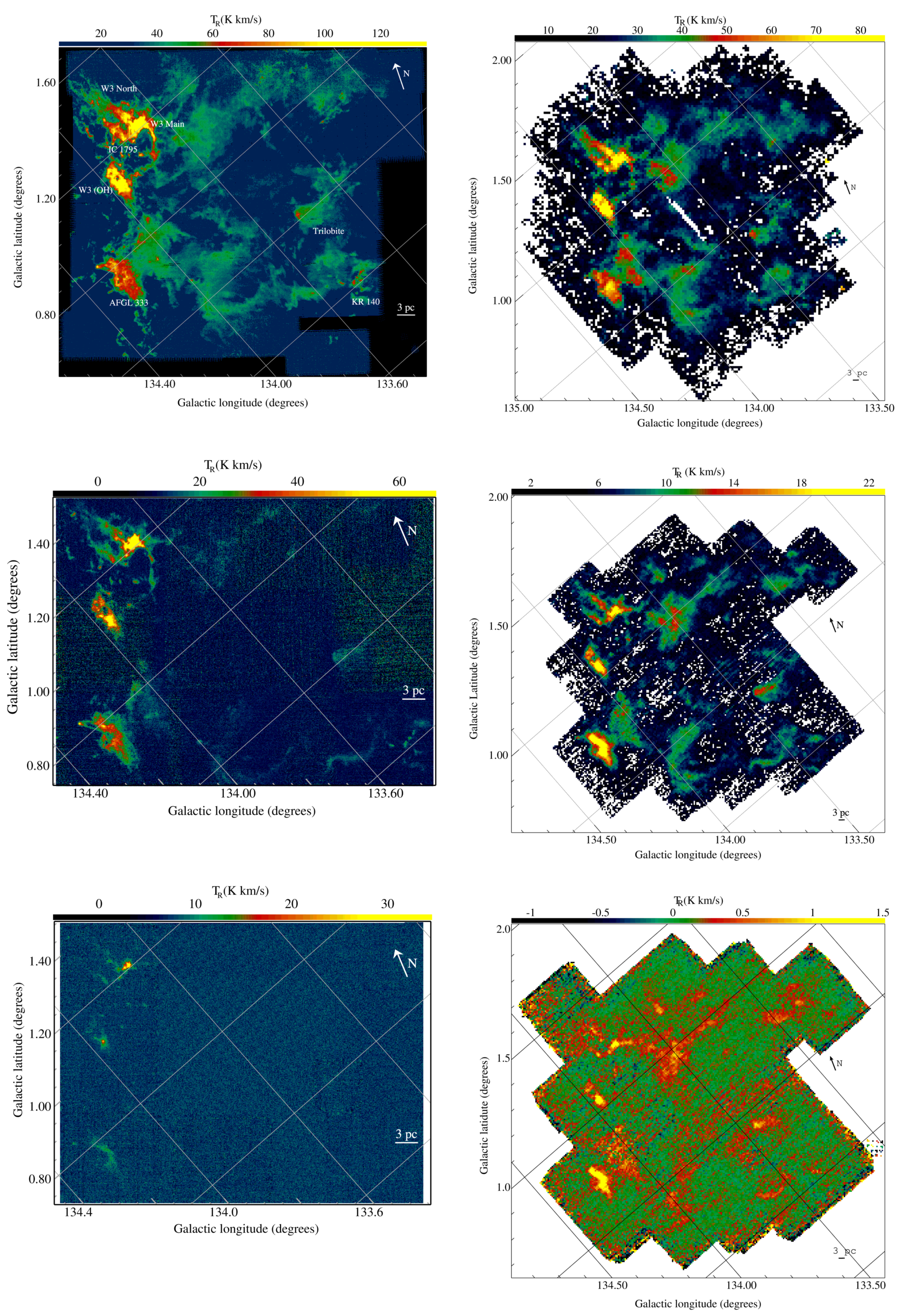}
\caption{The different isotope integrated T$_R$ (K\,km/s) emission line maps of the W3 GMC. {\bf Top Left:} $^{12}$CO J=3$\to$2; {\bf Top Right:} $^{12}$CO J=1$\to$0; {\bf Middle Left:} $^{13}$CO J=3$\to$2; {\bf Middle Right:} $^{13}$CO J=1$\to$0; {\bf Bottom Left:} C$^{18}$O J=3$\to$2; {\bf Bottom Right:} C$^{18}$O J=1$\to$0.}
\label{fig:coiso}
\end{figure*} 

The $^{12}$CO J=1$\to$0 observations at 115.271 GHz were made on 1999 May 7--14 and 2000 April 24--28 at the Five Colleges Radio Observatory (FCRAO) 14-m telescope using the 16-element SEQUOIA array receiver and FAAS backend with 256 channels and 80-MHz bandwidth giving 0.8-km\,s$^{-1}$ resolution.  System temperatures were in the range 600\,K to 800\,K producing rms noise of $T_A^{\star} \simeq 0.1$\,K.  the rms pointing correction was 5 arcseconds or less. These data and their reduction are more fully described in \citet{BrethertonPhD2003}.

The $^{13}$CO and C$^{18}$O J=1$\to$0 lines at 110.201 and 109.722 GHz, respectively, were observed simultaneously using the expanded 32-element SEQUOIA array on the FCRAO 14-m on 2004 March 16--23 March in ``On-The-Fly" continuous raster mode. The target region was covered with 32 individual submaps which are fully Nyquist sampled on to a 25$''$ grid with a spatial resolution of 49$''$. The spectrometer was used with 50-MHz bandwidth centred on $V_{\rm  LSR}= -40$\,km\,s$^{-1}$, which results in a velocity resolution of 133\,m\,s$^{-1}$. System temperatures were in the range $50-80$\,K for the duration of the observations. A fuller description of these data can be found in \citet{AllsoppPhD2011}.

All antenna temperatures have been changed to the $T_{\rm R}^{\star}$ scale using $T_{\rm R}^{\star} = T_{\rm A}^{\star} / \eta_{\rm fss}$ \citep{1981ApJ...250..341K}, where $\eta_{\rm fss}$ is the forward scattering and spillover efficiency, taken as 0.77 for HARP on JCMT \citep{2009MNRAS.399.1026B} and 0.70 for SEQUIOA on FCRAO \citep{1998ApJS..115..241H}.

\section[]{Results and analysis}
\label{sec:Analysis}
\subsection{The W3 GMC}

Emission from the W3 cloud was found in the range $-60 < V_{\mathrm{LSR}} < -30$\, \kms.  The velocity structure and dynamical state of the cloud will be discussed in detail elsewhere.  Figure \ref{fig:12co} shows the $^{12}$CO J=3$\to$2 emission integrated between --65\,\kms\ and --25\,\kms. These data cover the whole GMC, while the $^{13}$CO  and C$^{18}$O data in this transition cover a more limited area, slightly vignetting the northern and southern edges of the cloud.  The  integrated emission in these lines and in the three J=1$\to 0$ transitions are shown in Figure \ref{fig:coiso}. 

Figure \ref{fig:12co} shows that, in addition to the warm dense gas around the active star-forming regions, $^{12}$CO J=3$\to$2 also traces most of the diffuse, extended emission seen in lower-level transitions (Figure \ref{fig:coiso}, top right panel). This is somewhat surprising, since the critical density of the J=3$\to$2 transition is expected to be between $5 \times 10^4$\,cm$^{-3}$ at $T=40$\,K and $4\times10^5$\,cm$^{-3}$ at 10\,K \citep{1985MNRAS.214..271F}, slightly higher than that of CS J=1$\to$0, and the J=3 energy level is $E/ k=32.8$\,K above ground. The transition should thus trace the relatively warm, high-density gas associated with recent star formation. This may be explained by photon trapping caused by high optical depths which may be reduce the effective critical density.

Following \citet{1978ApJ...226L..39L} we identify as the high density layer the eastern edge of the cloud, adjacent to the bubble blown by the IC\,1805 OB cluster. The W3 North, W3 Main, IC\,1795, W3\,(OH) and AFGL\,333 regions are located here. For the purposes of our analysis, and consistency with Moore et al (2007), this is separated from the rest of the cloud (which we call the low density layer, LDL and includes the HB3, KR\,140 and the trilobite regions (see figure \ref{fig:msxw3w4})) by a line defined as $b = 1.2089 \times l + 162.7235$, in Galactic coordinates. This division is somewhat arbitrary and based on the visible extent of the intense star formation in the eastern portion of W3. However, while the definition of triggered and non-triggered cloud regions is not so clear cut, it is likely that the feedback effects from the W4 H{\sc ii} region will decrease in strength with distance from the IC\,1805 OB cluster. Given that the average integrated intensity in the HDL is three times higher than that of the LDL (23 K km/s compared to 7 K km/s), we assume, in this paper, that the expansion of the W4 Hii region has not yet affected these regions of W3. Therefore, star formation within the HDL, as defined by the equation above, is assumed to be triggered, whereas the LDL region is assumed to be dominated by spontaneous star formation. 

The three brightest and best-known star forming regions in the cloud (W3 Main, W3 (OH) and AFGL 333) are easily identified in Figure \ref{fig:12co}, running from north to south along the eastern edge of the cloud. W3 Main ($l=133.7095^{\rm o}$, $b=1.2500^{\rm o}$) is the most prominent of these and the brightest source in the region at many wavelengths. OH and H$_{2}$O masers have been detected towards sources IRS 4 and 5 (\citealp{1987ApJS...65..193G}; \citealp{1984ApJ...285L..79C}) and it is the richest \textsc{Hii} region cluster known within 2\,kpc from the Sun \citep{2008hsf1.book..264M}. We find a mean integrated intensity of $\int T_{\rm R}^{*} {\rm d}V$=307.5\,K$\,$km\,s$^{-1}$ with a peak of 700$\,$K$\,$km\,s$^{-1}$ in the central area and $\sim$55\,K\,km\,s$^{-1}$ in the more diffuse surrounding emission. The cloud associated with W3 Main is seen to have rather complex physical structure, possibly shaped by the cavity situated immediately to the south which has been created by the winds from the young IC\,1795 stellar cluster \citep{2008hsf1.book..264M}. 

On the other side of this cavity lies W3\,(OH) ($l=133.9515^{\rm o}$, $b=1.0600^{\rm o}$), another active high-mass star-forming region and also the host of OH and H$_{2}$O masers that point towards two centres of high mass star formation separated by 0.07\,pc (\citealp{1981MNRAS.195..213N}; \citealp{1977ApJ...215L.121F}). The molecular cloud associated with W3\,(OH) is clearly elongated north-south, which is consistent with it being part of the large-scale, compressed shell produced by the winds of the nearby IC\,1804 OB association responsible for the W4 \textsc{Hii} shell. Its brightest regions have a mean integrated $^{12}$CO J=3$\to 2$ intensity of $\int T_{\rm R}^{\star} \mathrm{d} \upsilon=167$\,K\,km\,s$^{-1}$ with a peak of 300\,K\,\kms, while the surrounding structure has a mean of $\sim$80\,K\,\kms. 

South of W3\,(OH) is located the third active star-forming region in the HDL, AFGL\,333 ($l=134.2030^{\rm o}$, $b=0.7630^{\rm o}$). This cloud has a less well defined central peak than either of W3\,Main or W3\,(OH) and has a mean integrated intensity of $49$\,K\,km\,s$^{-1}$. 

The rest of the W3 GMC contains less intense emission and less active star formation. On the south-eastern corner of the GMC, the cloud associated with the KR\,140 \textsc{Hii} bubble \citep{2008MNRAS.385..995K} is easily identifiable.

Between KR\,140 and AFGL\,333 is  a region we term Loops.  The CO emission in this area appears diffuse and generally has low integrated intensity ($\int T_R^{\star} \mathrm{d} \upsilon=15$\,K\,\kms).  In the 850-$\umu$m continuum, it appears as a long, fine, looped filament \citep{2007MNRAS.379..663M} and in Spitzer data is revealed to contain a string of infrared sources (Polychroni et al.\, in preparation). Above KR\,140 there is a region we call Trilobite. It has a mean integrated intensity of 14.4\,K\,\kms. Here, again, \citet{2007MNRAS.379..663M} find a number of dense cores at 850\,$\umu$m, indicative of star formation. While this region seems rather cut off from the rest of the cloud there is a bridge of diffuse material that connects it to the KR\,140 bubble. 

\begin{figure*}
\includegraphics[width=\textwidth]{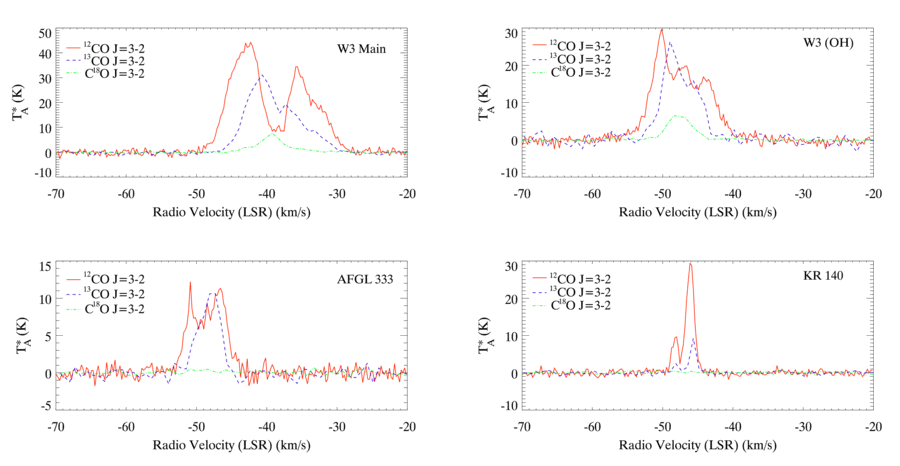}
\caption[small]{The $^{12}$CO, $^{13}$CO and C$^{18}$O J=3$\to$2 emission lines extracted from the cubes at $l=133.7095^{\rm o}$, $b=1.2149^{\rm o}$ (W3 Main), $l=133.9514^{\rm o}$, $b=1.0599^{\rm o}$ (W3\,(OH)), $l=133.2028^{\rm o}$, $b=1.7628^{\rm o}$ (AFGL\,333) and $l=133.4300^{\rm o}$, $b=1.4233^{\rm o}$ (KR\,140)}
\label{fig:isotopes}
\end{figure*}

In Figure \ref{fig:isotopes} we plot the emission lines of the three isotopes in J=3$\to2$ from individual pixels in the four main regions of the cloud (W3 Main, W3\,(OH), AFGL\,333 and KR\,140). It is clear from the spectra that the $^{12}$CO J=3$\to$2 emission line is optically thick and self-absorbed practically throughout the cloud. $^{13}$CO J=3$\to$2 also tends to be optically thick and self-absorbed, but only in the brightest regions like W3 Main or W3\,(OH). On the other hand C$^{18}$O J=3$\to$2 is very weak and we only observe it towards the brightest regions of the cloud (W3 Main, W3\,(OH) and AFGL\,333). 

Throughout this study we assume that the CO tracing the molecular gas of the W3 GMC is in a state of local thermodynamic equilibrium (LTE), at least in the rotation levels $J\le3$. Where the optical depth is high, as is the case for $^{12}$CO emission along most lines of sight, the effective critical density will be reduced due to photon trapping. This means that, in reality, there may be somewhat different critical densities for different isotopologues, and the safety of the LTE assumption may depend on the rarity of the CO species.  We ignore this in the following analysis but discuss it further below.

\begin{table}
\centering
\caption[The channel widths of $^{12}$CO, $^{13}$CO, C$^{18}$O J=1$\to$0 and J=3$\to$2 spectra.]{The channel widths of $^{12}$CO, $^{13}$CO, C$^{18}$O J=1$\to$0 and J=3$\to$2 spectra.}
\begin{tabular}{|c|c|c|}
\hline
Molecule & Transition & Channel Width (km\,s$^{-1}$) \\\hline
$^{12}$CO & 1$\to$0 & 0.8126 \\
$^{13}$CO &1$\to$0 & 0.1328 \\ 
C$^{18}$O &1$\to$0 & 0.1333 \\
$^{12}$CO &3$\to$2 & 0.2381 \\
$^{13}$CO &3$\to$2 & 0.8301 \\
C$^{18}$O &3$\to$2 & 0.3348 \\\hline
\end{tabular}
\label{tab:velchan}
\end{table}

The 3D cubes of all emission lines were collapsed along the velocity axis between --65\,\kms\ and --25\,\kms\ and multiplied by the velocity channel width (Table \ref{tab:velchan}) to produce integrated-intensity maps for each species and transition. The J=3$\to$2 data were re-gridded to match the J=1$\to$0 maps so that there was a one-to-one pixel correspondence.   

\subsection{Optical Depth}

The measured radiation temperature of a source is given, in terms of the excitation temperature $T_x$ and optical depth $\tau$, by the solution to the equation of radiation transfer, in the absence of a background source: 
\begin{equation}
\label{eqn:radtrans}
T_R=J\ (T_{x})\ \left(1-e^{-\tau}\right)
\end{equation}
where, in LTE,
\begin{equation}
\label{eqn:J}
J(T_{x})=\frac{h\nu}{k}\left(\frac{1}{e^{h\nu/kT_x}-1}-\frac{1}{e^{h\nu/kT_{\mathrm{bg}}}-1}\right),
\end{equation}
$\nu$ is the frequency and $T_{\mathrm{bg}}$ is the temperature of the cosmic microwave background (2.73K). Hence, if the same $T_x$ is assumed, the ratio of the line brightness temperatures in the same transition from two different isotopic species is given by
\begin{equation} 
\label{eq:od_1}
\frac{T_{\rm R,1}(j\to i)} {T_{\rm R,2}(j\to i)}= \frac{1-e^{-\tau}}{1-e^{-\tau/X}},
\end{equation}
where $\tau$ is the optical depth of the more abundant species and $X$ is the abundance ratio.

\begin{figure}
\centering
\includegraphics[height=6.cm]{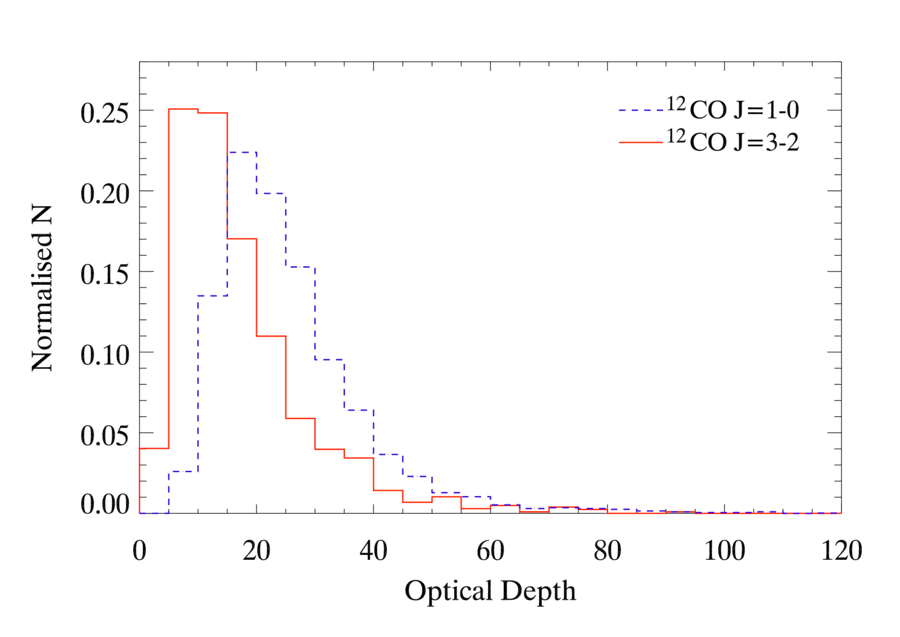}
\caption[small]{The optical depth distributions for $^{12}$CO J=3$\to$2 (red) and J=1$\to$0 (blue). For the $^{12}$CO J=3$\to$2 line we measure the mean=17, median=14 and the mode=11.2. For $^{12}$CO J=1$\to$0 we measure the mean=25, median=22 and the mode=10.8. }
\label{fig:tau_distr}
\end{figure}

We adopt a value of $X=77$ for the $^{12}$CO/$^{13}$CO abundance ratio \citep{2002A&A...390.1001S}. In this case, the results are not very sensitive to the choice of the value of $X$, particularly where the $^{12}$CO optical depth is high. Assuming $\tau(^{12}\mbox{CO}) \gg 1$, the numerator of equation \ref{eq:od_1} becomes approximately equal to unity, providing a first estimate for an iterative solution. We use this first estimate and the Newton-Raphson iterative method to solve equation \ref{eq:od_1} to calculate the velocity-averaged optical depth per pixel in $^{12}$CO and $^{13}$CO J=3$\to$2 and J=1$\to$0. We find that the range of $\tau(^{12}\mbox{CO})$ in both transitions is between 5 and 90 (see Figure \ref{fig:tau_distr}). At high temperatures we expect that $\tau_{32}(^{12}\mbox{CO})/\tau_{10}(^{12}\mbox{CO}) = 9$. However, globally, we find that this ratio is much lower, consistent with low gas temperatures.
$^{13}$CO emission is found to be optically thin across the cloud in both transitions, with the exception of a few pixels in J=3$\to$2 in the brightest regions (e.g.\ the central parts of W3 Main). 

\subsection{Excitation Temperature}
\label{subsec:Tx}

\begin{figure}
\centering
\includegraphics[width=8.4cm]{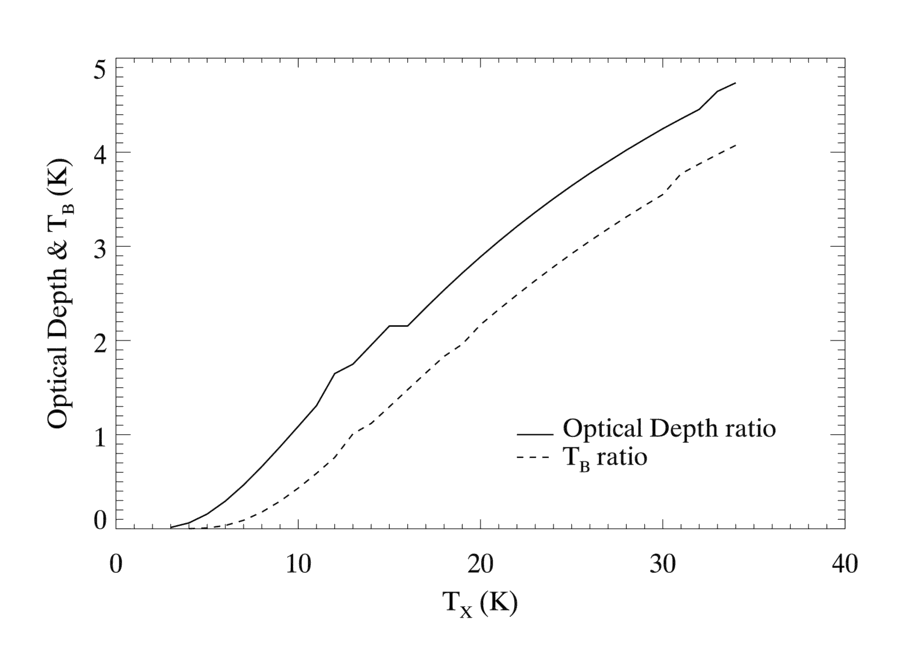}
\caption[small]{The solutions of the excitation temperature, T$_x$, as given from the $^{12}$CO J=3$\to$2 and J=1$\to$0 opacity ratios (solid line) and brightness temperature ratios (dashed line).}
\label{fig:txmodels}
\end{figure}

The excitation temperature $T_x$ parameterises the relative energy-level populations according to the Boltzmann distribution. Under the LTE assumption, $T_x$ is equal to the thermodynamic temperature of the gas.  Where optical depths  are determined for two transitions of the same species, the ratio of the two can be used to derive $T_x$, i.e.\ for the $^{12}$CO J=3$\to$2 and J=1$\to$0 transitions,
\begin{equation}
\label{eq:tau_rat}
\frac{\tau_{32}(^{12}\mbox{CO})}{\tau_{10}(^{12}\mbox{CO})}=3\ e^{-16.60/T_x} \ \frac{1-e^{-16.60/T_x}}{1-e^{-5.53/T_x}}.
\end{equation}
This recipe is obtained directly from equation \ref{eq:5}, assuming $\tau_{ji}$ is measured over the same velocity interval and distributed similarly over that interval for both transitions $j\to i$. Equation \ref{eq:tau_rat} cannot be solved analytically. Instead, we used it to compile a look-up table of optical depth ratio values as a function of $T_x$ in the range 3 to 34\,K (Figure \ref{fig:txmodels}). The look up table has a resolution of 0.5\,K. 

For those pixels in which complete optical depth information was not available, usually due to the inadequate detection of $^{13}$CO emission, we estimated $T_x$ from the ratio of line brightness temperatures, using a low optical depth approximation, as follows.

\begin{figure}
\includegraphics[height=6.7cm]{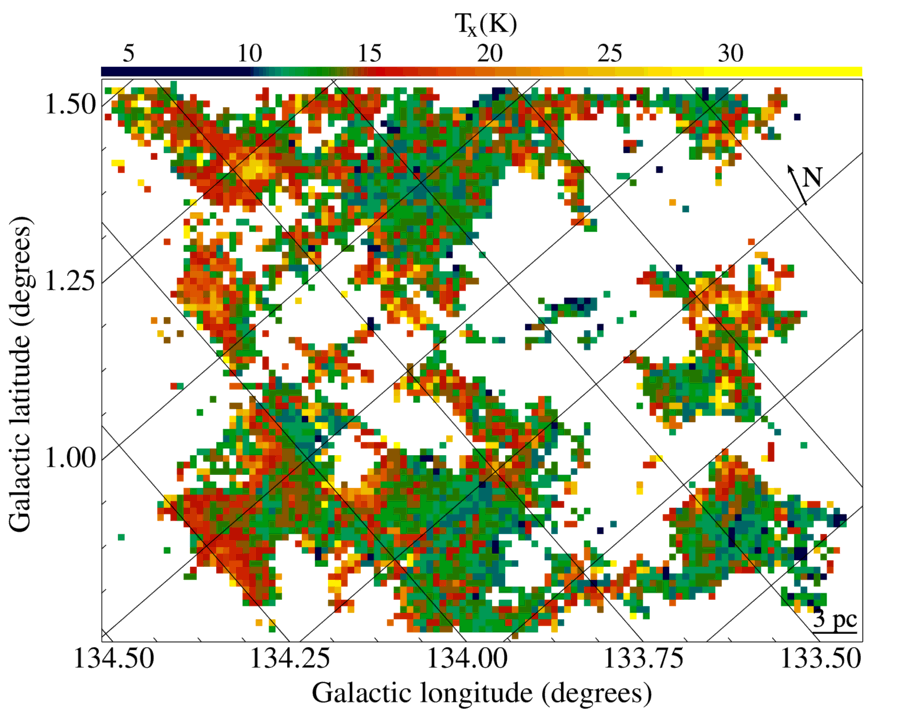}
\caption{The distribution of CO excitation temperature in the W3 GMC, extimated from the optical depth and integrated intensity ratios of $^{12}$CO J=3$\to$2 and J=1$\to$0. The axes are Galactic coordinates.}
\label{fig:Tx}
\end{figure}

Assuming $\tau \ll 1$, the ratio of the observed $^{12}$CO J=3$\to$2 and J=1$\to$0 line strengths is given by
\begin{equation}
\frac{T_{R,32}}{T_{R,10}}=3 \ \frac{\tau_{32}(^{12}CO)}{\tau_{10}(^{12}CO)} \ \frac{ \left( e^{16.60/T_x} - 1 \right)^{-1} - 2.29\times10^{-3}}{\left(e^{5.53/T_x}-1\right)^{-1}-0.152},
\label{eq:TR-rat}
\end{equation}
in which  $\tau_{32}(^{12}\mbox{CO})/\tau_{10}(^{12}\mbox{CO})$ is given by equation \ref{eq:tau_rat}. A look-up table was again used to obtain $T_x$ estimates from the observed line brightness ratios.  Figure \ref{fig:Tx} shows the distribution of derived excitation temperatures. The spatial resolution of the map is that of the J=1$\to$0 CO data, namely 44$''$. The temperature resolution is 1\,K, which is the resolution of the look-up table, and is less than the uncertainties in the data. Where $T_x >$\,10\,K this error is less than 10\%, while in regions of lower $T_x$), the error is around 20\%.

To estimate the error on the excitation temperature we calculated the average spectral noise per pixel for each transition. These noise maps were added to the integrated intensity maps and the calculation of $\tau$ and $T_x$ repeated. The error estimates quoted above are the average difference between the results in the nominal maps and those with added noise.

The majority of the CO-traced gas in the LDL has excitation temperatures around 8 -- 10\,K with small excursions to slightly higher values near star-forming regions (e.g. the trilobite and the KR 140 bubble).  The three main star-forming regions in the HDL all show significantly enhanced values of $T_x$ in the region of 15 -- 30\,K.  One feature to note is that, while the $T_x$ distributions in W3\,Main and W3\,(OH) tend to peak centrally, those in AFGL\,333 peak near the eastern edge of the cloud facing the W4 \textsc{Hii} region (Figure \ref{fig:Tx}). 

\subsection{Gas Mass Distribution}
\label{subsec:MCFE}

Having determined $T_x$ and $\tau$, the column density $N$ is obtained from equation \ref{eq:4}. For the $^{12}$CO J=3$\to$2 transition, and $\upsilon$ in \kms, the relationship is 

\begin{equation}
\label{eq:N32}
\frac{ N_{\mbox{\rm co}}}{\mathrm{m}^{-2}}= \frac{ 7.67 \times 10^{17} \left(T_x+0.922\right) e^{16.60/T_x} }{\left(1-e^{-16.60/T_x}\right) } \int \tau(\upsilon)d\upsilon. 
\end{equation}

\begin{figure}
\includegraphics[width=8.4cm]{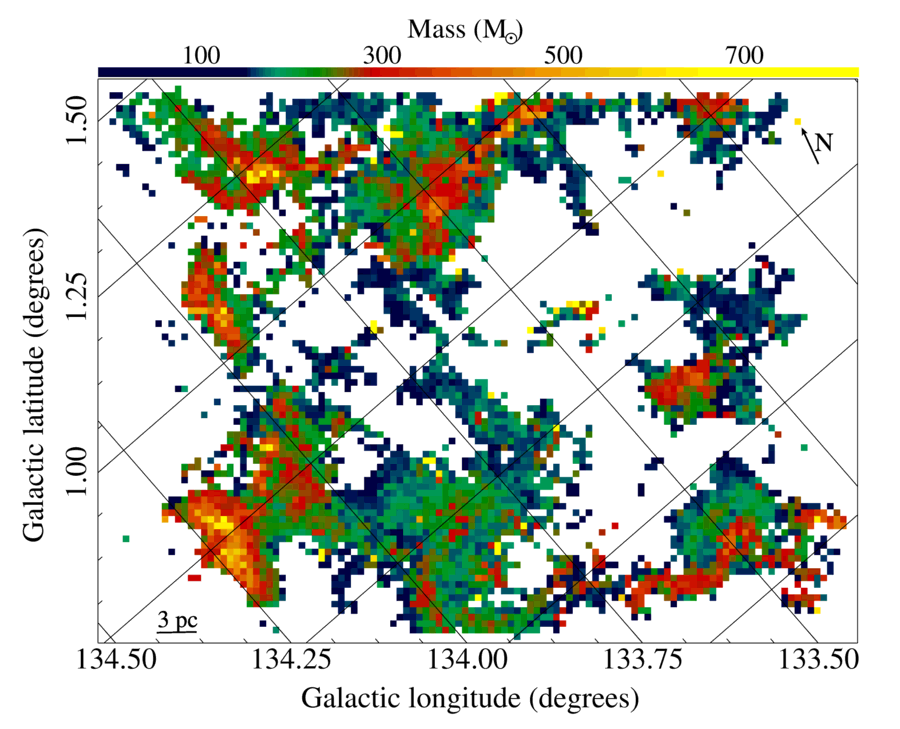}
\caption{The distribution of molecular gas mass (in $\msolar$ units) per pixel (0.43\,pc/pixel) across the W3 GMC. }
\label{fig:mass}
\end{figure}

\noindent
This is converted to molecular hydrogen column density using an abundance ratio of $[^{12}\mbox{CO}]/[\mbox{H}_2]=9.8 \times 10^{-5}$ \citep{1982ApJ...262..590F} and to mass per pixel (Figure \ref{fig:mass}) assuming a distance to the cloud of 2\,kpc (\citealp{2006Sci...311...54X}; \citealp{2006ApJ...645..337H}). Integrating over the map, we find that the W3 GMC has a total mass of $4.4\pm0.4 \times 10^{5}\msolar$, consistent with previous estimates of the cloud's mass. The error on the mass was estimated in the same way as for $T_x$, as described above in Section \ref{subsec:Tx}. 

The accuracy of the absolute values in Figure \ref{fig:mass} is limited by the uncertainty in the value of [$^{12}$CO]/[H$_{2}$].  The value we use is derived from extinction measurements of the nearby Taurus and $\rho$ Ophiuchi molecular cloud \citep{1982ApJ...262..590F}. Direct measurements of this abundance (e.g. \citealp{1994ApJ...428L..69L}; \citealp{1985ApJ...298..316W}) can differ by factors of 3 to 5 which makes other sources of error, propagated from the $\tau$ and $T_x$ calculations, insignificant.  However, this is a systematic error, and does not affect comparisons made within the map. In terms of random error, the most well-behaved regions are those with the highest optical depth ($\tau>$\,20), where uncertainties are of the order $\pm$10$\msolar$ per pixel. In regions with lower excitation temperatures ($T_x<$\,10), for which the uncertainties in the optical depth produce higher relative errors, the uncertainty on the mass per pixel is, on average, $\pm$20 $\msolar$. This translates to an error of as low as 4\% in pixels with mass higher than $\sim$ 250\,$\msolar$ and as high as 30\% in pixels with mass lower than $\sim$ 60$\msolar$.

The spatial distribution of mass shown in Figure \ref{fig:mass} closely follows that of the $^{12}$CO J=3$\to$2 emission in Figure \ref{fig:12co}, as well as of the 850-$\umu$m continuum in the HDL and southern part of the cloud \citep{2007MNRAS.379..663M}.  

\section{Discussion}
\label{sec:discussion}

The CO J=3$\to$2 emission line data are similar in mapping extent to the existing FCRAO CO J=1$\to$0 data as well as the SCUBA continuum observations of the cloud \citep{2007MNRAS.379..663M}. The data contain a very large amount of detailed information on the physical state of the W3 GMC.  This paper concentrates on the distribution of gas temperatures inferred from LTE excitation temperatures and the results of subsequent calculations of the distribution of mass in the cloud.  The velocity structure and dynamics of the cloud will be the subject of a subsequent paper.
 
\subsection{Gas temperatures}

\subsubsection{$T_x$ method}

Our excitation temperature results are obtained by assuming that LTE conditions apply throughout the GMC, i.e.\ that rotational levels $J\le3$ are thermalised, populated according to the Boltzmann distribution which is dependent only on temperature.  There are several caveats to this assumption.   The first is that the critical density of the 3$\to$2 transition is quite high, as mentioned above, and it is quite likely that the mean density in the majority of the gas comprising W3 is less than this.  Where this is the case, the energy-level populations will be determined by the collision rate, and so by both the temperature and density of the gas.  $T_x$ will then underestimate the kinetic temperature $T$ of the gas and the observed line radiation temperature $T_{\rm R}$ will be less than predicted by LTE.  The effect on predicted column densities is complex.  Whether or not $N$ is under- or over-estimated depends at least partly on the gas temperature.  At low $T$, an underestimate of $T$ may cause an overestimate of $N$ and at high $T$ the reverse may be true.  At temperatures around 30\,K small errors in $T$ will have little effect.  A second point to note is that the effective critical density of a transition may be lower where the optical depth is high and photon trapping becomes significant.  This is likely to apply to the $^{12}$CO transitions across most of the cloud and to $^{13}$CO along high column density lines of sight and in the line centres.  This effect may undermine the assumption of equal $T_x$ for the same transition of different isotopic species used to calculate optical depths.

These issues notwithstanding, our method of deriving $T_x$ from the ratio of the optical depths of two different transitions of the same species, or from the line radiation temperature ratio where optical depths are low, has some advantages.  Firstly, the value of $\tau_{32}/\tau_{10}$ depends on the populations of all four energy levels and so the derived value of $T_x$ represents the distribution of energies relatively well (being equivalent to a fit of the Boltzmann distribution) and should give a fairly robust estimate of column density, even if $T_x$ underestimates the real kinetic temperature of the gas.  Secondly, equation \ref{eqn:radtrans} contains, implicitly, the filling factor ($\eta_{\rm \,ff}$) of the emission within the telescope beam, i.e.\ the fraction of the beam area filled by the emitting gas.  Although $\eta_{\rm \,ff}$ may not be quite the same for different transitions, it will be accounted for, to first order, by using $T_{\rm R}$ and $\tau$ ratios.  

In other studies, $T_x$ has often been estimated from the brightness temperature of a single transition, using equation \ref{eqn:radtrans} (e.g.\ \citealt{2010MNRAS.401..204B}).  This is the best method in the absence of data in other transitions, but the LTE assumption then models only the relative populations of the upper and lower levels of the one transition used (J=3 and J=2 in this case) and does not account for $\eta_{\rm \,ff}$, which may be quite small.

\subsubsection{$T_x$ results}

The large-scale distribution of $T_x$ revealed in Figure \ref{fig:Tx} contains few surprises.  In general, we see higher temperatures ($>$20\,K) near regions of active star formation and cooler gas ($\le$10\,K) elsewhere.  We also see two large-scale temperature gradients, one running east-west across the whole cloud and the other along a north-south axis through the HDL. Along the first axis, temperatures range from $\sim$20-30\,K in the HDL down to $\sim$4-9\,K in the central and western regions of the GMC. It is well known (e.g. \citealp{2007IAUS..237..481U}) that in regions of high density the gas is in thermal equilibrium with the radiatively heated dust, whereas at lower densities molecular cooling dominates and the gas cools down through molecular transitions. Therefore, this gradient is likely to be the result of lower densities to the west of the HDL as well as a gradient in the radiation field intensity. The second gradient ranges from $\sim$30\,K in W3 Main to $\sim$10\,K in AFGL 333. Since gas temperatures may be an indicator of the evolutionary stage of star formation within a cloud, this north-south trend may imply an age sequence. Such an age gradient has also been suggested by \citet{2005JKAS...38..257S} who observed all three regions in atomic carbon. 

In addition to this, there are some detailed differences in the $T_x$ distribution within the three bright HDL regions. In W3 Main, $T_x$ peaks clearly in the middle of the associated cloud, coincident with the brightest IR and submm sources. The embedded YSOs in W3 Main therefore appear to be the dominant heating source in the cloud.  The surrounding molecular gas may, in fact, be in the process of being dispersed by these centrally formed objects.  W3\,(OH) has a lower average excitation temperature of about 20\,K. $T_x$ is also centrally peaked in this source, although less clearly than in W3 Main, and there are also high temperatures along its western edge, indicating that external heating may be important in this cloud. Finally, AFGL\,333 exhibits the lowest mean excitation temperature ($\sim$10\,K) of these three regions and the $T_x$ distribution clearly peaks at the eastern edge, which is exposed to the radiation from the IC\,1805 cluster (Figure \ref{fig:msxw4w3}). Assuming the embedded YSOs become more dominant heating sources with time, these internal $T_x$ distributions appear to support the idea of an age gradient from north to south along the HDL. 

\subsection{Mass Distribution}

Obtaining the distribution of $T_x$ has allowed us to derive the mass distribution of the cloud with much more accuracy than previous studies which assume a single temperature. Our new estimate of the total mass of the GMC is $(4.4\pm0.4) \times 10^{5}\,\msolar$, consistent with that of 
\citet{2007MNRAS.379..663M} who obtained $(3.8 \pm 1.1) \times 10^{5}\,\msolar$ from $^{13}$CO J=1$\to$0 data, assuming $T_x = 30$\,K everywhere.  This agreement is despite most of the cloud having $T_x < 30$\,K which, because we are below $E/k = 33$\,K, should produce higher mass estimates. We find that the mass is almost equally divided between the HDL region and the remainder of the cloud ($2.23 \times 10^{5}\,\msolar$ and $2.19 \times 10^{5}\,\msolar$, respectively), even though the latter covers almost twice as much projected area as the HDL.

\subsubsection{Clump Formation Efficiency}

\begin{figure}
\includegraphics[height=7.2cm]{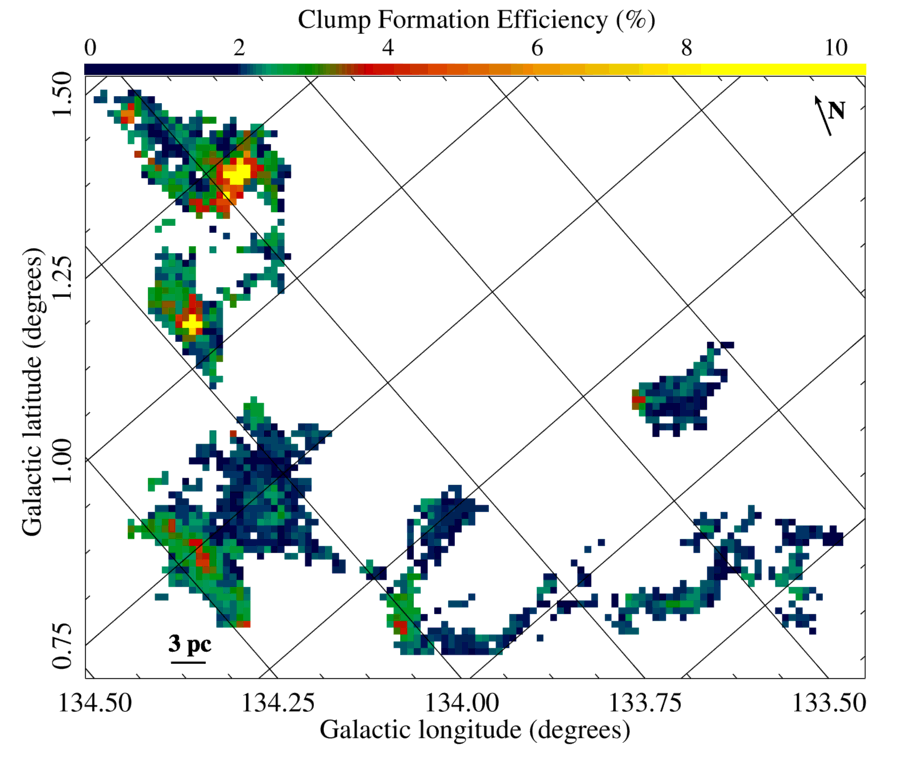}
\caption{The clump formation efficiency (\%) as a function of position across the W3 GMC. The CFE is given by the ratio of the sub-millimetre mass and the CO-traced molecular gas mass, calculated where the 850\,$\mu$m continuum data was available.}
\label{fig:cfe}
\end{figure}

We use the existing SCUBA observations of the cloud \citep{2007MNRAS.379..663M} along with the masses derived above to calculate the fraction of gas in dense, potentially star-forming  structures as a function of position in the cloud, i.e.\ the clump formation efficiency (CFE). The CFE is a time-integrated quantity and can be written as:
\begin{equation}
\mbox{CFE}=\frac{1}{M_{\rm cloud}} \int_{t=0}^{t=\mbox{\tiny now}} \dot{M}(t) \,{\rm d}t,
\end{equation}
where $\dot{M}$ is the rate of formation of dense-core mass from the available gas of the cloud. Therefore a high CFE can be the result of either a high average dense-core formation rate or of a long integration time.  We have calculated the CFE for the area of W3 surveyed at 850$\umu$m by Moore et al (2007). Sub-millimetre flux densities are converted to gas mass using the standard formula,
\begin{equation}
 M=\frac{S_{\nu}D^{2}}{\kappa_{\nu}B_{\nu}(T_{d})}
\end{equation}
where $S_{\nu}$ is the integrated flux density at 850\,$\umu$m, $D$ is the distance to the cloud, $\kappa_{\nu}$ is the mass absorption coefficient and B$_{\nu}(T)$ is the Planck function. $T_d$ is the dust temperature, for which we assume a constant value of 20\,K, representative of the dust temperatures of dense cores derived from infrared SEDs  (e.g. \citealp{2010A&A...518L..97E}; \citealp{2010A&A...520L...8P}). We assume a constant value of $\kappa_{\nu}$=0.01\,cm$^{2}$g$^{-1}$ \citep{2001ApJ...556..215M} and so a constant gas-to-dust ratio throughout the cloud. The SCUBA map is then regridded to match the CO-traced mass data. The division of these two maps gives the distribution of the CFE across the cloud (Figure \ref{fig:cfe}). 

As was found for $T_x$, there are two large-scale CFE gradients across the GMC.  One runs north-south along the HDL where W3 Main with CFE$\sim$23\% and W3\,(OH) ($\sim$20\%) exhibit higher efficiency in gas-to-clump conversion than AFGL\,333 ($\sim$3\%).  In comparison, the spontaneous star-forming regions in the western section of W3 have very low values of CFE, generally below 1\%, creating a second, east-west gradient. The most obvious interpretation is that the high CFE values present in the eastern regions of the cloud, are the direct result of the interaction between with the adjacent expanding W4 \textsc{Hii} superbubble. This is agreement with the results of \citet{2007MNRAS.379..663M} who found that 26\% of the gas has been converted into dense clumps with $M \ge 13$\,M$_{\odot}$ in the HDL, compared to 5\% in the western regions of the cloud. They also found that the Core Mass Function (CMF) of the cloud does not change significantly between the triggered and spontaneous star-forming regions of the cloud. There is clear evidence, therefore, that triggering due to the W\,4 expansion results in the creation of new dense structure (clumps/cores) across the affected regions that is able to form new stars. The common CMF of the two regions suggests that the triggering does not affect the star-formation process any further than that, i.e. triggering allows the formation of more more massive stars, but it does not alter the shape of the cannonical IMF.

This is in agreement with the collect-and-collapse model \citep{1994A&A...290..421W}, where a trigger, such as the winds from massive stars, can create a shock that propagates through the surrounding medium, sweeping and compressing the gas adjacent to the forming bubble, creating new dense structure that can become gravitationally unstable along its surface on long timescales. 

The higher efficiencies measured in W3 Main, compared to the W3\,(OH) and AFGL\,333 regions, can also be interpreted as due to different timescales.  Assuming that the triggering from the W4 bubble affects the three regions in a similar way, i.e. it increases the star formation rate to a similar degree, then the difference of a factor of three in the clump formation efficiency between W3 Main and AFGL\,333 star-forming regions may be simply due to W3 Main being older than AFGL\,333. This is consistent with the results of several studies that indicate multiple generations of star formation in and around W3 Main (e.g. \citealp{2008ApJ...673..354F}; \citealp{2011ApJ...743...39R}).
  
\subsection{Mach Number}

The turbulent fragmentation model of star formation neatly provides a mechanism for the simultaneous support of molecular clouds against gravity on large scales and the formation of dense, star-forming cores in the collisions between turbulent flows on small scales, thus naturally explaining the low star-formation efficiency (SFE) usually found \citep{2007ARA&A..45..565M}.  A relation should therefore exist between turbulence and the CFE in a given cloud. \citet{2009arXiv0907.0248P} use the \citet{2005IAUS..227..276M} star-formation rate (SFR) model, extended to a magnetised medium, to study the relationship between the SFR, the virial parameter, $\alpha_{vir}$, and the sonic rms Mach number. Higher Mach numbers mean stronger shocks, which should produce thicker, denser compressed regions as well as additional support against gravity on large scales. The relationship, therefore, is not a simple one.  However, the models predict a weak negative correlation between the star formation rate and the Mach number for turbulence-dominated star-forming regions. This implies that there should be more large-scale support against gravity where the turbulence is stronger, as expected.

In order to investigate this prediction, we have measured the velocity width, $\sigma_{co},$ of the J=3$\to$2 emission. Unfortunately, the emission line widths are very much dependent on whether the line is self-absorbed and/or optically thick. Thus, in this analysis we cannot use the $^{12}$CO transition since it is largely both self-absorbed and optically thick. $^{13}$CO also suffers from optical depth effects towards the densest and brightest regions of the cloud and we, therefore, have to be very careful when using it to derive the velocity widths. Generally, C$^{18}$O is the best choice as it is optically thin, however, it is a weak line and we only detect it in the HDL regions of the cloud. For the $^{13}$CO J=3$\to$2 we find that the line widths vary between 0.6 and 4.0\,km/s while for the C$^{18}$O J=3$\to$2 emission line they vary between 0.5 and 3.0\,km/s.   

The total velocity dispersion in the molecular gas can be obtained from $\sigma_{co}$ by deconvolving the thermal velocity dispersion of the CO molecules, estimated from the sound speed $\sqrt{3kT/m_{\rm co}}$, and convolving the thermal velocity dispersion of the mean molecular gas, i.e.\ the sound speed $c_s = \sqrt{3kT/\mu m_{\rm H}}$. Here $\mu$ is the mean molecular weight = 2.8, assuming solar neighbourhood abundances. The gas temperature $T$ can be estimated using the CO excitation temperature $T_x$ obtained above (where $T_x = T_{kin}$ assuming LTE). This results in $c_s$=0.2-1.3\,km\,s$^{-1}$ through the cloud which is systematically lower than the line widths calculated above for $^{13}$CO and C$^{18}$O, confirming the presence of supersonic flows. If the velocity distribution is assumed to be Gaussian, we can express the total three-dimensional velocity dispersion in the CO-traced gas as

\begin{equation}
\label{eq:M2}
\sigma^{2}_{3D}(\mathrm{total})=
3\left[\sigma^{2}_{\rm co}-\frac{kT}{m_{\rm H}} \left( \frac{1}{\mu_{\rm co}}-\frac{1}{\mu} \right) \right]
\end{equation}
where $\mu_{\rm co} = 29$ and 30 for $^{13}$CO and C$^{18}$O, respectively.
\noindent
The Mach number, $\mathcal{M}$, is given by:
\begin{equation}
\mathcal{M}^2=\frac{\sigma^2_{3D}(\mathrm{total})}{c^2_s}.
\end{equation}
Using equation \ref{eq:M2}, this can be written
\begin{equation}
\mathcal{M}^{2}= 3.4\times10^{-4} \, \frac{\sigma^{2}_{co}}{T} - x
\end{equation}
where $x = 0.903$ for $^{13}$CO and 0.907 for C$^{18}$O.

\begin{figure}
\includegraphics[height=5.3cm]{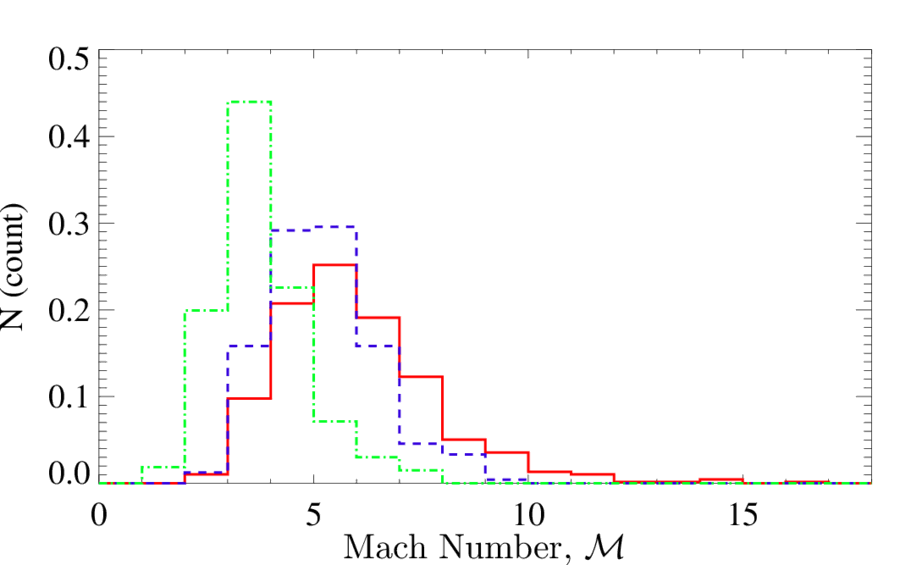}
\caption{The $\mathcal{M}$ normalised distributions for the HDL region as derived from $^{13}$CO (red line) and C$^{18}$O J=3$\to$2 (green dashed-dot line) and for the LDL region as derived from $^{13}$CO J=3$\to$2 (blue dashed line).}
\label{fig:mdist}
\end{figure}

Figure \ref{fig:mdist} show the range of $\mathcal{M}$ values in the LDL as derived from $^{13}$CO (blue dashed line) and in the HDL from $^{13}$CO (red line) and C$^{18}$O (green dash-dot line). From the $^{13}$CO it is clear that while there are higher values of $\mathcal{M}$ in the HDL region the distribution does not differ significantly from that of the LDL. This is due to the effects of optical depth as well as higher noise in the more diffuse gas that the $^{13}$CO traces in bot the HDL and LDL regions. The C$^{18}$O-derived MAch number distribution, on the other hand, does not suffer from optical depth effects and suffers less from noise levels as it comes from the brightest regions. Therefore it should be more indicative of the real $\mathcal{M}$ distribution in the densest regions.

\begin{figure}
\includegraphics[height=11.5cm]{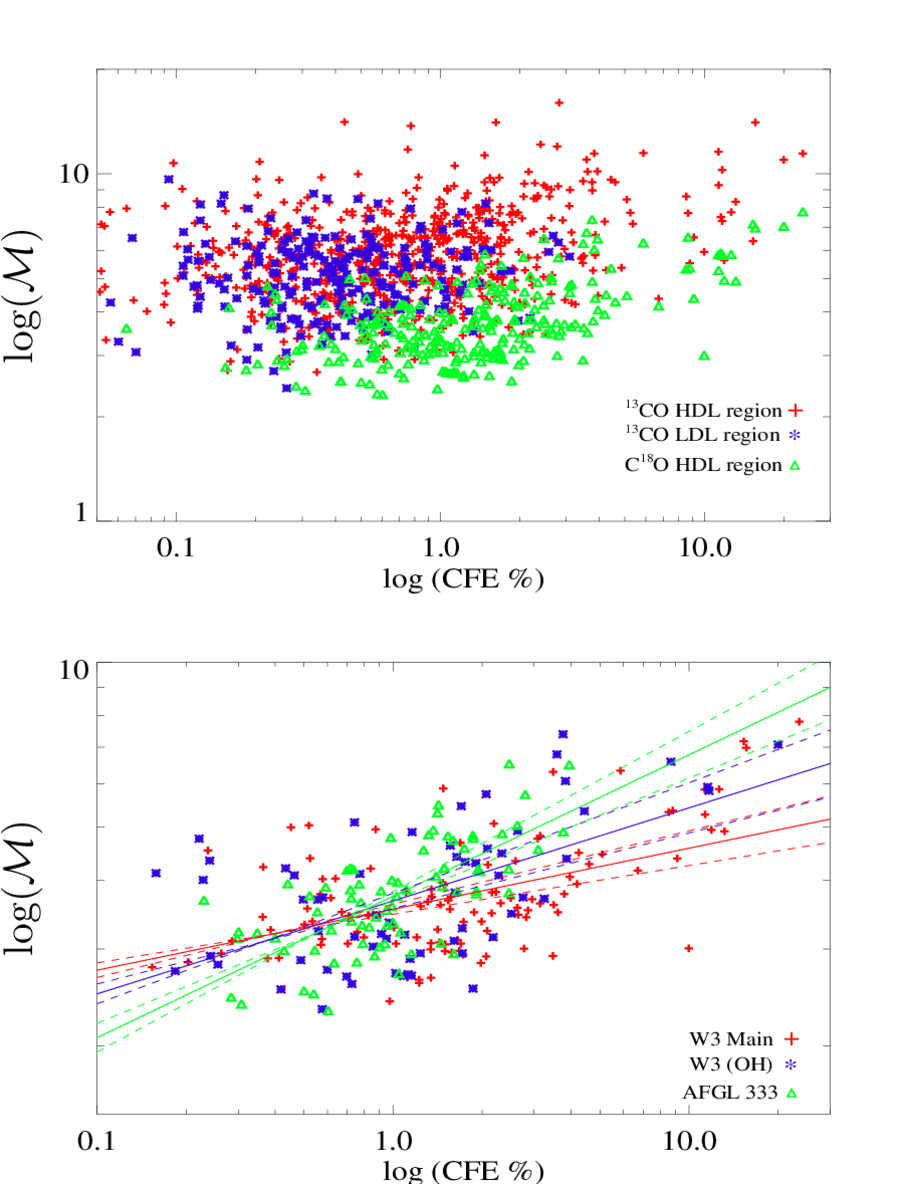}
\caption{{\bf Top:} $\mathcal{M}$ per pixel plotted against the Clump Formation Efficiency in the HDL (red crosses for the $^{13}$CO derived $\mathcal{M}$ and green triangles for the C$^{18}$O derived $\mathcal{M}$) and LDL (blue stars for the $^{13}$CO derived $\mathcal{M}$). {\bf Bottom:} $\mathcal{M}$ derived from the C$^{18}$ J=3$\to$2 emission line plotted against the CFE across the three HDL regions. The fitted lines are the linear least-square fits to the three regions. The gradients and intercepts for the three regions are: m=0.11$\pm$0.02 and $\beta$=0.55$\pm$0.01 for W3 Main, m=0.17$\pm$0.03 and $\beta$=0.56$\pm$0.01 for W3\,(OH) and m=0.26$\pm$0.03 and $\beta$=0.57$\pm$0.01 for AFGL\,333. The dashed lines indicate the 1-$\sigma$ confidence interval in the fit.}
\label{fig:mach2p}
\end{figure}

The first panel of figure \ref{fig:mach2p} shows $\mathcal{M}$ plotted against CFE for the LDL region from $^{13}$CO J=3$\to$2 (blue stars) and the HDL region from both $^{13}$CO (red crosses) and C$^{18}$O J=3$\to$2 (green triangles). Note that we have included the $\mathcal{M}$ calculated from all the pixels in all three maps, where there is signal, i.e. there are two velocity dispersion estimates for regions that have signal in both the $^{13}$CO and C$^{18}$O maps. Comparisons between the HDL and LDL regions are only possible in the $^{13}$CO as we have practically no detections in the LDL in C$^{18}$O. To determine whether there is a correlation between $\mathcal{M}$ and CFE we used the Spearman Rank correlation test for all the subsets. For the $^{13}$CO data in the HDL we find that despite the spread there is positive correlation significant at the 3 sigma level ($\rho$=0.25, t=6.7, N=675). In the LDL region, on the other hand we find that there is no correlation between $\mathcal{M}$ and the CFE ($\rho$=0.066, t=1.2, N=240). The spread in the $^{13}$CO J=3$\to$2 derived $\mathcal{M}$ is because of the more diffuse material that surround the densest regions in W3 Main, W3 (OH) and AFGL 333 triggered regions. To minimise this effect and also to negate the effect introduced due to the $^{13}$CO lines being optically thick in the densest regions, we use the C$^{18}$O line, as it only traces the most dense regions and is also optically thin. We find a tighter correlation between $\mathcal{M}$ and the CFE significant at a 3 sigma level ($\rho$=0.47, t=850, N=261).
 
We therefore find no evidence in the current data to support the prediction of \citet{2009arXiv0907.0248P}. It is possible that the lack of correlation between $\mathcal{M}$ and CFE in the western regions of the cloud, which should be dominated by spontaneous star formation, could be due to the small range in CFE values found there. The significant positive correlation found in the HDL can have two interpretations. It could be the result of the expansion shock produced by the W4 shell interacting with the gas in these regions, increasing the turbulent velocities and the efficiency with which the cloud has formed dense structure, in a way that is inconsistent with the predictions of turbulent (spontaneous) star formation. A more likely interpretation is that the increased degree of turbulence is an effect of the higher CFE present in these regions, coupled with their age, i.e.\ it is due to feedback from the on-going star formation injecting momentum or kinetic energy, which results in higher measured $\mathcal{M}$. 

In the bottom panel of figure \ref{fig:mach2p} we again plot $\mathcal{M}$, as derived from the C$^{18}$O J=3$\to$2 emission line against CFE per pixel, but separate the data into the three main HDL star-forming regions, W3 Main (red crosses), W3\,(OH) (blue stars) and AFGL\,333 (green triangles), revealing that the relationship between these quantities is different for each region. Spearman rank correlation tests on the three subsamples produce the results $\rho$=0.47, 0.51 and 0.65 with t= 5.6, 4.06, 7.4, and N=115, 63, 77 data points respectively, which are all significant at a level $>3 \sigma$. 

A least-square fit to each of the three regions gives different results for each of them with gradient and intercept values of m=0.11$\pm$0.02 and $\beta$=0.55$\pm$0.01 for W3 Main, m=0.17$\pm$0.03 and $\beta$=0.56$\pm$0.01 for W3\,(OH) and m=0.26$\pm$0.03 and $\beta$=0.57$\pm$0.01 for AFGL\,333. We note, further, that while there is only a small difference of 1-2 $\sigma$ between the slopes of W3 Main/W3\,(OH) and of W3\,(OH)/AFGL\,333, there is a larger difference of 3-4 $\sigma$ between the slopes of W3 Main and AFGL\,333.

This change in the steepness of the slopes in the above correlations for the three different regions can be related to the evolutionary stage and state of each. If the observed turbulence were simply due to the mechanical interaction with the W4 expansion, and the CFE were determined by the turbulence, then all three regions should have more or less the same correlation between $\mathcal{M}$ and the CFE. The change in the correlation between the three regions can be interpreted as a result of the on-going star formation affecting the cloud. Feedback processes, like outflows or stellar winds from forming stars, have been acting on these regions for different lengths of time. In W3 Main the slope of the correlation is flatter than in W3\,(OH), which again is less steep than that found in AFGL\,333. Since W3 Main is thought to be more evolved than W3\,(OH) and AFGL\,333 it follows that the flattening of the slope is likely to be because of CFE values produced by a longer star-formation timescale. 

\section{Conclusions}
\label{sec:conc}

We have presented new CO J=3$\to$2 maps of the W3 giant molecular cloud, obtained using HARP on the JCMT.  In conjunction with FCRAO CO J=1$\to$0 data, we have used these maps to derive the gas properties as a function of position within the W3 GMC. 

We have used the ratio of the optical depths of the two transitions of $^{12}$CO (where $\tau\,>1$) and the ratio of the brightness temperatures T$_{b}$ (where $\tau\,<1$) to derive the distribution of excitation temperature in the CO-traced molecular gas (Figure \ref{fig:Tx}). We find high excitation temperatures (T$_{x}>12\,K$) in the eastern HDL region, where there is active star formation. In the remainder of the GMC the temperature rarely rises above 10\,K. We see a temperature gradient along the HDL, where star formation has been triggered by compression due to expansion of the nearby W4 H{\sc ii} region. We associate this with an age gradient in which W3 Main is the most evolved of the main star-forming regions, followed by W3\,(OH) and AFGL\,333. 

Using the excitation temperature map, we have obtained an accurate determination of the distribution of gas mass in the W3 GMC (Figure \ref{fig:mass}). We find that the cloud contains $4.4 \pm 0.36 \times 10^{5}\msolar$, half of which is located in the HDL region. This value is in agreement with previous estimates ($3.8 \pm 1.1 \times 10^{5}\msolar$; \citet{2007MNRAS.379..663M}).

We used existing sub-millimetre continuum observations of the cloud \citep{2007MNRAS.379..663M} to measure the so-called clump-formation efficiency (CFE) i.e.\ the fraction of molecular gas in the form of dense, potentially star-forming structures (clumps), as a function of position in the cloud. We find that, in the regions affected by the expanding W4 \textsc{Hii} superbubble, the CFE has values of 3-25\%, much higher than that of the rest of the cloud that remains apparently unaffected, where values are less than $\sim$1\%. We conclude that the triggering mechanism that has created the actively star-forming HDL in the W3 GMC primarily works by creating new dense structures, in agreement with the collect-and-collapse model \citep{1994A&A...290..421W}, rather than by forcing the collapse of existing structures in the gas.

We have used the widths of the $^{13}$CO and C$^{18}$O J=3$\to$2 emission lines to derive the sonic rms Mach Number across the GMC. We find that there is a positive correlation between the Mach Number and the CFE, but only in the gas of the HDL. This correlation is opposite to that expected from models of turbulence-driven star formation and is probably due to feedback from the recent star formation injecting momentum into the nearby gas. The slope of this correlation is different (Figure \ref{fig:mach2p}) in each of the three main star-forming region in this part of the cloud (i.e. W3 Main, W3 (OH) and AFGL 333) and we interpret this as another indicator of the differing evolutionary stages of these three regions.

\section{Acknowledgements}

We would like to thank the reviewer, Chris Davis, for his comments on this paper as they helped make it better. We would also like to acknowledge Eugenio Schisano and Diego Turrini for useful discussions relating to this work and Tim Jennes for his help with the JCMT HARP data reduction. DP wishes to acknowledge a STFC PhD studentship for this work. The James Clerk Maxwell Telescope is operated by The Joint Astronomy Centre on behalf of the Science and Technology Facilities Council of the United Kingdom, the Netherlands Organisation for Scientific Research, and the National Research Council of Canada. The data were obtained under Program IDs m06bu21, m07bu17 and m08bu24. FCRAO was supported by NFS Grant AST 08-38222. This research has made use of NASA's Astrophysics Data System.

\onecolumn

\appendix
\section{Derivation of LTE formulae}
\label{sec:app}

The absorption coefficient of a transition between energy levels $j$ and $i$ at frequency $\nu_{ji}$ is

\begin{equation}
\kappa(\nu_{ji}) = \frac{h \nu_{ji}}{4\pi} \left( n_i B_{ij} - n_j B_{ji} \right) \phi(\nu_{ji}).
\end{equation}
Hence, with the relative energy level populations $n_j/n_i$ populated according to the Boltzmann distribution under LTE and with the normal relations between the Einstein coefficients, we have

\begin{equation}
\kappa(\nu_{ji}) = \frac{c^2}{8\pi} \frac{g_i}{g_j} n_i \frac{A_{ji}}{\nu^2_{ji}}
\left( 1 - e^{-h \nu_{ji} / k T_x} \right) \phi(\nu_{ji}).
\end{equation}
Then, since $n_i/n$ is also determined by Boltzmann, 

\begin{equation}
\kappa(\nu_{ji}) = \frac{c^2}{8\pi} g_j n \frac{1}{\mathcal{Z}} e^{-h \nu_{i0}/kT_x} 
\frac{A_{ji}}{\nu^2_{ji}}
\left( 1 - e^{-h \nu_{ji} / k T_x} \right) \phi(\nu_{ji})
\end{equation}
where $\mathcal{Z}$ is the partition function.  Now since $\tau(\nu) = \int \kappa(\nu) \,
\mathrm{d}s \simeq \kappa(\nu) L$, where $L$ is the optical path, and $g_j = 2j+1$,

\begin{equation}
\tau(\nu_{ji}) = \frac{c^2}{8\pi} (2j+1) \mathcal{Z} e^{-h \nu_{i0}/kT_x} 
\frac{A_{ji}}{\nu^2_{ji}}
\left( 1 - e^{-h \nu_{ji} / k T_x} \right) \phi(\nu_{ji}) N_{ji}(\nu)
\end{equation}

\noindent
Then, integrating over the line, the column density of molecules emitting in the $j\to i$ transition is given by:

\begin{equation}
\label{eq:1}
N_{ji}=\frac{8\pi}{c^2(2j+1)}\ \mathcal{Z} \ e^{\frac{h\nu_{i0}}{kT_{x}}}\ \frac{\mathrm{\nu}_{ji}^2}{A_{ji}}\ \left(1-e^{-\frac{h\nu_{ji}}{kT_{x}}}\right)^{-1} \int\tau(\nu) \ \mathrm{d} \nu
\end{equation}
where $\mathcal{Z}$ is the partition function and $h \nu_{i0}$ is the energy of level $i$ above ground.  The Einstein $A$ coefficient is given by:

\begin{equation}
\label{eq:2}
A_{ji}=\frac{16\pi^{3}\mathrm{\nu}_{ji}^3}{3\epsilon_{o}hc^3}\,\mu_{ji}^{2} \, \frac{j}{2j+1}
\end{equation}
where $\mu$ is the dipole moment of the molecule.  Substituting this in \ref{eq:1} gives:
\begin{equation}
\label{eq:3}
N_{ji}=\frac{3\epsilon_{o}hc}{2\pi^{2}\nu_{ji}\mu^{2}j} \ \mathcal{Z} \ e^{\frac{h\nu_{i0}}{kT_{x}}} \ \left(1-e^{-\frac{h\nu_{ji}}{kT_{x}}}\right)^{-1}\int\tau(\nu) \ \mathrm{d} \nu
\end{equation}

\noindent
The partition function, $\mathcal{Z}$, can be approximated empirically by:
\begin{equation}
\label{eq:4}
\mathcal{Z}=\frac{k}{h B}\left (T_x+\frac{h B}{3 k}\right).
\end{equation}

Hence, for CO with $\mu = 0.112$\,Debye and replacing the integral over frequency with an integral over velocity $\upsilon$, 

\begin{equation}
\label{eq:5}
N_{ji}=\frac{2.30 \times 10^{15}}{j} \ \left(T_x + 0.922 \right) \ e^{\frac{h\nu_{i0}}
{kT_{x}}} \ \left(1-e^{-\frac{h\nu_{ji}}{kT_{x}}}\right)^{-1}\int\tau(\upsilon) \ \mathrm{d} \upsilon
\end{equation}

\label{lastpage}


\begin{thebibliography}{99}

\bibitem[\protect\citeauthoryear{Allsopp} {2011}]{AllsoppPhD2011} Allsopp J., 2011, PhD thesis, Liverpool John Moores University.

\bibitem[\protect\citeauthoryear{Bertoldi}{1989}]{1989ApJ...346..735B} Bertoldi F., 1989, ApJ, 346, 735 

\bibitem[\protect\citeauthoryear{Bertoldi \& McKee}{1990}]{1990ApJ...354..529B} Bertoldi F., McKee C.~F., 1990, ApJ, 354, 529 

\bibitem[\protect\citeauthoryear{Bretherton} {2003}]{BrethertonPhD2003} Bretherton D.E., 2003, PhD thesis, Liverpool John Moores University.

\bibitem[\protect\citeauthoryear{Buckle et al.}{2010}]{2010MNRAS.401..204B} 
Buckle J.~V., et al., 2010, MNRAS, 401, 204 

\bibitem[\protect\citeauthoryear{Buckle et al.}{2009}]{2009MNRAS.399.1026B} 
Buckle J.~V., et al., 2009, MNRAS, 399, 1026 

\bibitem[\protect\citeauthoryear{Claussen et 
al.}{1984}]{1984ApJ...285L..79C} Claussen M.~J., et al., 1984, ApJ, 285, 
L79 

\bibitem[\protect\citeauthoryear{Deharveng, Zavagno, \& Caplan}{2005}]{2005A&A...433..565D} Deharveng L., Zavagno A., Caplan J., 2005, A\&A, 433, 565 

\bibitem[\protect\citeauthoryear{Elia et 
al.}{2010}]{2010A&A...518L..97E} Elia D., et al., 2010, A\&A, 518, L97 

\bibitem[\protect\citeauthoryear{Elmegreen \& Lada}{1977}]{1977ApJ...214..725E} Elmegreen B.~G., Lada C.~J., 1977, ApJ, 214, 725 

\bibitem[\protect\citeauthoryear{Elmegreen}{1998}]{1998ASPC..148..150E} Elmegreen B.~G., 1998, ASPC, 148, 150 

\bibitem[\protect\citeauthoryear{Feigelson 
\& Townsley}{2008}]{2008ApJ...673..354F} Feigelson E.~D., Townsley L.~K., 2008, ApJ, 673, 354 

\bibitem[\protect\citeauthoryear{Flower 
\& Launay}{1985}]{1985MNRAS.214..271F} Flower D.~R., Launay J.~M., 1985, MNRAS, 214, 271 

\bibitem[\protect\citeauthoryear{Forster, Welch, 
\& Wright}{1977}]{1977ApJ...215L.121F} Forster J.~R., Welch W.~J., Wright M.~C.~H., 1977, ApJ, 215, L121 

\bibitem[\protect\citeauthoryear{Frerking, Langer, 
\& Wilson}{1982}]{1982ApJ...262..590F} Frerking M.~A., Langer W.~D., Wilson R.~W., 1982, ApJ, 262, 590 

\bibitem[\protect\citeauthoryear{Gaume 
\& Mutel}{1987}]{1987ApJS...65..193G} Gaume R.~A., Mutel R.~L., 1987, ApJS, 65, 193 

\bibitem[\protect\citeauthoryear{Ginsburg, Bally, \& Williams}{2011}]{2011MNRAS.418.2121G} Ginsburg A., Bally J., Williams J.~P., 2011, MNRAS, 418, 2121 

\bibitem[\protect\citeauthoryear{Hachisuka et 
al.}{2006}]{2006ApJ...645..337H} Hachisuka K., et al., 2006, ApJ, 645, 337

\bibitem[\protect\citeauthoryear{Heitsch et 
al.}{2006}]{2006ApJ...648.1052H} Heitsch F., Slyz A.~D., Devriendt 
J.~E.~G., Hartmann L.~W., Burkert A., 2006, ApJ, 648, 1052 


\bibitem[\protect\citeauthoryear{Heyer et al.}{1998}]{1998ApJS..115..241H} 
Heyer M.~H., Brunt C., Snell R.~L., Howe J.~E., Schloerb F.~P., Carpenter 
J.~M., 1998, ApJS, 115, 241 

\bibitem[\protect\citeauthoryear{Kerton et al.}{2008}]{2008MNRAS.385..995K} 
Kerton C.~R., Arvidsson K., Knee L.~B.~G., Brunt C., 2008, MNRAS, 385, 995 


\bibitem[\protect\citeauthoryear{Klein, Sandford, 
\& Whitaker}{1980}]{1980SSRv...27..275K} Klein R.~I., Sandford M.~T., II, Whitaker R.~W., 1980, SSRv, 27, 275 


\bibitem[\protect\citeauthoryear{Klessen et 
al.}{2004}]{2004ASPC..322..299K} Klessen R.~S., Ballesteros-Paredes J., Li 
Y., Mac Low M.-M., 2004, ASPC, 322, 299 


\bibitem[\protect\citeauthoryear{Kutner 
\& Ulich}{1981}]{1981ApJ...250..341K} Kutner M.~L., Ulich B.~L., 1981, ApJ, 250, 341 


\bibitem[\protect\citeauthoryear{Lacy et al.}{1994}]{1994ApJ...428L..69L} 
Lacy J.~H., Knacke R., Geballe T.~R., Tokunaga A.~T., 1994, ApJ, 428, L69 


\bibitem[\protect\citeauthoryear{Lada et al.}{1978}]{1978ApJ...226L..39L} 
Lada C.~J., Elmegreen B.~G., Cong H.-I., Thaddeus P., 1978, ApJ, 226, L39 


\bibitem[\protect\citeauthoryear{Mac Low 
\& Klessen}{2004}]{2004RvMP...76..125M} Mac Low M.-M., Klessen R.~S., 2004, RvMP, 76, 125 


\bibitem[\protect\citeauthoryear{McKee 
\& Chakrabarti}{2005}]{2005IAUS..227..276M} McKee C.~F., Chakrabarti S., 2005, IAUS, 227, 276 


\bibitem[\protect\citeauthoryear{McKee 
\& Ostriker}{2007}]{2007ARA&A..45..565M} McKee C.~F., Ostriker E.~C., 2007, ARA\&A, 45, 565 


\bibitem[\protect\citeauthoryear{Megeath et 
al.}{2008}]{2008hsf1.book..264M} Megeath S.~T., Townsley L.~K., Oey M.~S., 
Tieftrunk A.~R., 2008, hsf1.book, 264 


\bibitem[\protect\citeauthoryear{Mitchell et 
al.}{2001}]{2001ApJ...556..215M} Mitchell G.~F., Johnstone D., 
Moriarty-Schieven G., Fich M., Tothill N.~F.~H., 2001, ApJ, 556, 215 


\bibitem[\protect\citeauthoryear{Moore et al.}{2007}]{2007MNRAS.379..663M} 
Moore T.~J.~T., Bretherton D.~E., Fujiyoshi T., Ridge N.~A., Allsopp J., 
Hoare M.~G., Lumsden S.~L., Richer J.~S., 2007, MNRAS, 379, 663 

\bibitem[\protect\citeauthoryear{Norris 
\& Booth}{1981}]{1981MNRAS.195..213N} Norris R.~P., Booth R.~S., 1981, MNRAS, 195, 213 

\bibitem[\protect\citeauthoryear{Oey 
\& Clarke}{2005}]{2005ApJ...620L..43O} Oey M.~S., Clarke C.~J., 2005, ApJ, 620, L43 


\bibitem[\protect\citeauthoryear{Padoan 
\& Nordlund}{2002}]{2002ApJ...576..870P} Padoan P., Nordlund {\AA}., 2002, ApJ, 576, 870 


\bibitem[\protect\citeauthoryear{Padoan 
\& Nordlund}{2009}]{2009arXiv0907.0248P} Padoan P., Nordlund A., 2009, arXiv, arXiv:0907.0248 

\bibitem[\protect\citeauthoryear{Paradis et al.}{2010}]{2010A&A...520L...8P} Paradis D., et al., 2010, A\&A, 520, L8 

\bibitem[\protect\citeauthoryear{Rivera-Ingraham et al.}{2011}]{2011ApJ...743...39R} Rivera-Ingraham A., Martin P.~G., Polychroni D., Moore T.~J.~T., 2011, ApJ, 743, 39 

\bibitem[\protect\citeauthoryear{Sakai, Oka, 
\& Yamamoto}{2005}]{2005JKAS...38..257S} Sakai T., Oka T., Yamamoto S., 2005, JKAS, 38, 257 


\bibitem[\protect\citeauthoryear{Sch{\"o}ier et 
al.}{2002}]{2002A&A...390.1001S} Sch{\"o}ier F.~L., J{\o}rgensen J.~K., van Dishoeck E.~F., Blake G.~A., 2002, A\&A, 390, 1001 


\bibitem[\protect\citeauthoryear{Sugitani et 
al.}{1989}]{1989ApJ...342L..87S} Sugitani K., Fukui Y., Mizuni A., Ohashi 
N., 1989, ApJ, 342, L87 

\bibitem[\protect\citeauthoryear{Urban 
\& Evans}{2007}]{2007IAUS..237..481U} Urban A., Evans N.~J., 2007, IAUS, 237, 481 

\bibitem[\protect\citeauthoryear{Watson et al.}{1985}]{1985ApJ...298..316W} 
Watson D.~M., Genzel R., Townes C.~H., Storey J.~W.~V., 1985, ApJ, 298, 316 


\bibitem[\protect\citeauthoryear{Whitworth et 
al.}{1994}]{1994A&A...290..421W} Whitworth A.~P., Bhattal A.~S., Chapman S.~J., Disney M.~J., Turner J.~A., 1994, A\&A, 290, 421 


\bibitem[\protect\citeauthoryear{Whitworth et 
al.}{1994}]{1994MNRAS.268..291W} Whitworth A.~P., Bhattal A.~S., Chapman 
S.~J., Disney M.~J., Turner J.~A., 1994, MNRAS, 268, 291 

\bibitem[\protect\citeauthoryear{Xu et al.}{2006}]{2006Sci...311...54X} Xu 
Y., Reid M.~J., Zheng X.~W., Menten K.~M., 2006, Sci, 311, 54 

\end{thebibliography}
\end{document}